\documentstyle[11pt,chpaspconf,epsf,psfig,twoside]{article}
\begin{document}
\def\mbold#1{\mbox{\boldmath$#1$\unboldmath}}
\def\etal{et\,al.}
\def\sqig{$\sim\,$}
\def\sun{$_{\scriptstyle\odot}$}
\def\msun{$M_{\scriptstyle\odot}$}
\def\swave{$S$-wave}
\def\chisq{$\chi^{2}_{\nu}$}
\def\minone{$^{-1}$}
\def\mintwo{$^{-2}$}
\def\up#1{$^{\mbox{{\scriptsize #1}}}$}
\def\lo#1{$_{\mbox{{\scriptsize #1}}}$}
\def\sqiglt{\hbox{\rlap{\lower.55ex \hbox {$\sim$}}
	\kern-.3em \raise.4ex \hbox{$<$}\,}}
\def\sqiggt{\hbox{\rlap{\lower.55ex \hbox {$\sim$}}
	\kern-.3em \raise.4ex \hbox{$>$}\,}}
\def\phiorb{$\phi_{\mbox{{\scriptsize orb}}}$}
\def\phispin{$\phi_{\mbox{{\scriptsize spin}}}$}
\def\pten#1{$\times10^{#1}$}
\def\deg{$^{\circ}$}
\def\cps{counts s$^{-1}$}
\def\EXO{{\sl EXOSAT}}
\def\asca{{\sl ASCA}}
\def\rosat{{\it Rosat}}
\def\ginga{{\it Ginga}}
\def\einstein{{\it Einstein}}
\def\beat{$\omega$\,--\,$\Omega$}
\def\ofr{(\kern-0.1em r\kern-0.1em )}
\def\gev{\,Ge\kern-.10em V}
\def\mev{\,Me\kern-.10em V}
\def\kev{\,ke\kern-.10em V}
\def\ev{\,e\kern-.10em V}
\def\kmps{km\,s$^{-1}$}
\def\fourpieps{4\pi\epsilon_{0}}
\def\ditto{\tt "}
\def\Ha{H$\alpha$}
\def\Hb{H$\beta$}
\def\Hg{H$\gamma$}
\def\Hd{H$\delta$}
\def\HeI{He\,{\sc i}}
\def\HeII{He\,{\sc ii}}
\def\HeIIl{He\,{\sc ii}\,$\lambda$4686}
\def\HeIl{He\,{\sc i}\,$\lambda$4471}
\def\HeIs{He\,{\sc i}\,$\lambda$6678}
\def\la{$\lambda$\,}
\renewcommand{\textfraction}{0.0}
\renewcommand{\topfraction}{1.0}

\setcounter{page}{1}


\title{Recent results on intermediate polars}
\author{Coel Hellier}
\affil{Department of Physics, Keele University, Staffordshire ST5 5BG, U.K.}

\begin{abstract}
I review recent activity in the field of intermediate polars,
concentrating on: the mode of accretion (disc-fed, disc-overflow
or discless); accretion curtains (the transition region and the
accretion footprint); X-ray pulse profiles (occultation and
absorption effects); accretion columns (mass determinations,
line-broadening, the soft X-ray component), and outbursts
in intermediate polars.   
\end{abstract}

\section{Introduction}
The intermediate polars (IPs) have been reviewed comprehensively by Patterson
(1994) and Warner (1995), see also Hellier (1995; 1996). Here I 
discuss areas of current activity, mainly from the observational
side, and with an X-ray bias. A good place to start is with a
(fairly conservative) census of currently known systems, presented on
the now traditional spin--orbit diagram (Fig.~1).  Compared
to previous versions we have a tripling
of the known systems below the gap, with RX\,1238--38 (Buckley \etal\ 1998)
and RX\,0757+63 (Tovmassian \etal\ 1998) joining EX~Hya.

\section{Accretion Mode} 
Almost all IPs show a dominant X-ray periodicity at the spin
period (a defining characteristic). Some, though, show 
additional X-ray periodicities, particularly at the beat frequency \beat\ 
(where $\omega$\,=\,spin frequency and
$\Omega$\,=\,orbital frequency). This was seen first in TX~Col (Buckley \&\
Tuohy 1989) and subsequently in at least 5 others
(e.g.\ Hellier 1991; 1998). The now-standard interpretation is that 
some of the accretion stream overflows the initial impact with the disc and 
connects directly onto field lines, producing `flipping' 
between the poles on the beat period (e.g.\ Hellier 1991). This has
become known as `disc-overflow' or `stream-overflow' accretion. It appears
to be variable, judging by the changing ratio of the 
spin and beat amplitudes in stars such as FO~Aqr and TX~Col (Hellier 1991; 
Norton \etal\ 1997; Wheatley, this volume).  

\begin{figure}[t]   
\vspace*{7.5cm}
\includegraphics{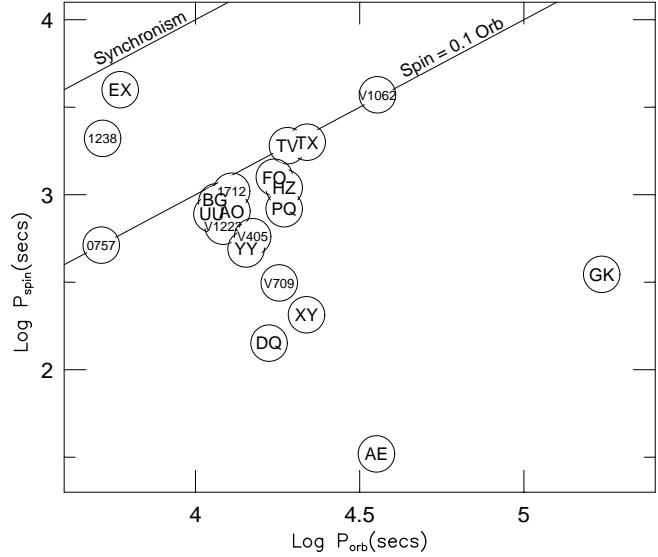}
\caption{The updated spin--orbit diagram for intermediate polars.}
\end{figure}

\begin{table}[t]
\begin{center}\begin{small}
\begin{tabular}{lcccclccc} \\ [-6mm] 
Star\rule{0mm}{4.3mm} & P\lo{spin}\,(s)\hspace*{-2mm} & 
P\lo{orb}\,(h)\hspace*{-2mm} & m\lo{v} & \rule{1mm}{0mm}  &
Star & P\lo{spin}\,(s)\hspace*{-2mm} & P\lo{orb}\,(h)\hspace*{-2mm} & m\lo{v}  \\ [1mm] \hline  
EX~Hya\rule{0mm}{4.3mm} & 4022 & 1.638 & 13 & & PQ~Gem & 833 & 5.19 & 14.5  \\
V1062~Tau\hspace*{-3mm} & 3700: & 9.96: & 15.1 & & AO~Psc & 805 & 3.59 & 13.5  \\ 
RX\,1238--38\hspace*{-3mm} & 2147 & 1.41:  & 16 &&  V1223~Sgr & 746 & 3.37 & 13 \\
TX~Col & 1911 & 5.72 & 15 && V405 Aur & 545 & 4.15 & 14.6  \\
TV~Col & 1910 & 5.49 & 14 && RX\,0757+63\hspace*{-3mm} & 515 & 1.44 & 16 \\  
WX~Pyx & 1558 &  & 17  && YY~Dra & 529 & 3.96 & 16  \\
RX\,0153+74\hspace*{-3mm} & 1414 &  &  && GK~Per & 351 & 47.9 & 14  \\
FO~Aqr & 1254 & 4.85 & 13.5 & & V709~Cas & 313 & 5: & 14.6 \\
HZ~Pup & 1210 & 5.1 & 18.5 && XY~Ari & 206 & 6.06 & $>$23 \\
RX\,1712--24\hspace*{-3mm} & 927 & 3.4: & 14 && DQ~Her & 142 & 4.65 & 15 \\
BG~CMi & 913  & 3.23 & 14.5  && AE~Aqr & 33 & 9.88 & 11  \\
UU~Col & 864 & 3.45 & 17.6 & \\ [1mm] 
\hline 
\end{tabular} \\ [2mm] \end{small}\parbox[t]{12.5cm}{{\small
References in Patterson 1994, and Hellier \etal\ 1998 
(RX\,1238--38); Still \etal\ 1998 (V405~Aur); Tovmassian \etal\ 1998 
(RX\,0757+63); O'Donoghue \etal\ 1996 (WX~Pyx); Motch \etal\ 1996 
(RX\,0153+74, V709 Cas); unpublished (HZ~Pup); Buckley \etal\ 1997
(RX\,1712--24); Burwitz \etal\ 1996 (UU~Col).\vspace{-6mm}}} 
\end{center}\end{table}

This idea was criticised by Murray at this conference using the following 
argument.  The overflowing
stream can't penetrate further in than the ballistic distance of
minimum approach to the white dwarf, $r_{\rm min}$. If the disc
disruption, $r_{\rm mag}$,  occurs inside $r_{\rm min}$, the overflowing
stream would re-impact on the disc, not the magnetosphere. If, though,
$r_{\rm mag} > r_{\rm min}$, it is unclear whether a disc can form. 
Murray \etal\ (1999, see also this volume) propose that spiral 
shocks could provide an alternative means by which material at the 
magnetosphere retains knowledge of orbital phase, and hence produces 
sideband periodicities. 

My preference is still for the disc-overflow model, because in
X-ray lightcurves the dominant sideband periodicity (where seen)
is at \beat, rather than 2(\beat) (e.g.\ Hellier 1992).  The \beat\
`pole-flipping' frequency arises naturally when two poles are fed from
one location in orbital phase, as from a stream (e.g.\ Hellier 1991;
Wynn \&\ King 1992). A twin-armed spiral leads most readily to twice
this frequency, or 2(\beat) (Murray \etal\ 1999, and this volume). 

So can we overcome Murray's argument?  In addressing it we
encounter a far more general problem for IP accretion (see also Warner
1996). I'll use FO~Aqr, often showing
a \beat\ modulation (Hellier 1993; Beardmore \etal\ 1998), as a test
case. Firstly, FO~Aqr is almost certainly close to equilibrium rotation,
since the $O$\,--\,$C$ of the spin cycle has alternated between spin-up and spin-down
over the last decade (Patterson \etal\ 1998). We can thus assume that the 1254-s spin
period is also the Keplerian period at the inner edge of the disc
(i.e.\ a critical fastness parameter \sqig 1). Using the standard
formulae for CV parameters (as collected in Warner 1995), with 
$P_{\rm orb}$\,=\,4.85\,hr and assuming $M_{1} = 0.7$\,M$_{\odot}$, we find 
that as a fraction of the stellar separation, $r_{\rm min}$\,=\,6\%, the
stream's circularization radius, $r_{\rm circ}$\,=\,10\%, and 
$r_{\rm mag}$\,=\,14\%.  

This is good news for the
disc-overflow model, in that the stream would hit the magnetosphere.
Indeed, this is insensitive to $q$ and 
applies to most IPs.\footnote{Only in GK~Per, with its long 2-d
orbit, would $r_{\rm min}$ be well outside $r_{\rm mag}$, and here
Hellier \&\ Livio (1994) proposed that a re-impact on the disc
resulted in the X-ray QPO seen in outburst
(though see Morales-Rueda, Still \&\ Roche, this volume,
for a contrary opinion).}
However, having $r_{\rm mag}$\,$>$\,$r_{\rm circ}$ is bad news for the
disc (its angular momentum would dissipate).
Since IPs do seem to be predominantly disc-fed (e.g.\ the dominance of
X-ray pulsations at $\omega$; Hellier 1991), this is a general problem
for the theory of magnetic accretion and points to the need
for torques not currently taken into account, either produced by the
disc-field interaction, or by the magnetic interaction of the two stars
(e.g.\ Warner 1996). Note that Mason (1997), studying the large spin-down
rate in PQ~Gem, suggested that material was threading onto field
lines well outside the co-rotation radius, in contradiction to the
standard theory (e.g.\ Ghosh \&\ Lamb 1979). 

If we rule out $r_{\rm mag}$\,$>$\,$r_{\rm circ}$, the disc-overflow
model can still work in the regime
$r_{\rm circ}$\,\sqiggt $r_{\rm mag}$\,\sqiggt $r_{\rm min}$ provided
that the disc can form. Whether this is so is unclear, though if 
the centrifugal barrier of a spinning magnetosphere makes accretion 
inefficient, a build up of material could screen enough of the field
to allow a disc to form. Lamb \& Melia (1988) suggest that this always 
occurs. 

Lastly, note that $r_{\rm mag}$ is set by the dense disc.  If an
overflowing stream is relatively tenuous (Mukai, Ishida \&\ Osborne 1994 
estimated that it carried only
2\%\ of the accretion flow in FO~Aqr) it could be penetrated by the field
further out than the disc is, allowing stream--field interactions
even with $r_{\rm mag} <  r_{\rm min}$.

The above arguments also lead to the expectation that faster rotators, 
which have smaller magnetospheres, would be less likely to show 
disc-overflow sidebands; Norton \etal\ (1999) suggest that this is indeed
the case.

\subsection{Optical sidebands}
As a result of the disc-overflow idea becoming popular to explain
X-ray sidebands, some recent papers have used it to explain
optical sidebands also. While this is possible, it should be remembered
that  there is nothing wrong with the 
traditional appeal to reprocessing of spin-pulsed X-rays by the secondary
and/or hot-spot. Indeed, detailed studies of the phasing of the 
optical beat period often support this (e.g.\ Hellier, Cropper \&\
Mason 1991).  The reason that this won't work in the X-ray is that
the combination of the solid-angle of the target and the albedo to
X-ray reflection means that the amplitude can be at most a few
percent of that of the spin-pulsed X-rays. (See also Wickramasinghe
\&\ Ferrario, this volume, for an account of optical sidebands.)

One thing that has proved puzzling is the occasional observation
of optical pulsations at 2(\beat), with little or nothing seen at
\beat\ (e.g.\ in TV~Col and RX\,1238--38; Buckley \&\ Sullivan 1992;
Buckley \etal\ 1998).
A twin-armed spiral shock could well explain this 
(e.g.\ Murray \etal\ 1999) since reprocessing off diametrically-opposite
sites would double the frequency and so give 2(\beat) rather than \beat.

\subsection{Stream-fed accretion}
Through Buckley \etal's papers (1995, 1997) we know of one system, 
 RX\,1712--24, that appears to accrete solely by a stream, since \beat, rather
than $\omega$, dominates the X-ray lightcurves. 
Fig.~2 shows unpublished H$\beta$ profiles folded on the \beat\ cycle, which 
appear to show the stream
flipping in velocity between the poles, as expected in a discless accretor.
We heard relatively little
about RX\,1712--24 at this conference, partly because the tumbling
of the magnetosphere beneath the stream produces chaotic variability
that is hard to analyse. For instance a Fourier transform of an
{\sl ASCA\/} lightcurve (unpublished) shows significant power only at 
2(\beat), in contrast to the \beat\ detected by Buckley \etal\ (1997). 

\begin{figure}[t]       
\vspace*{7.6cm}
\includegraphics{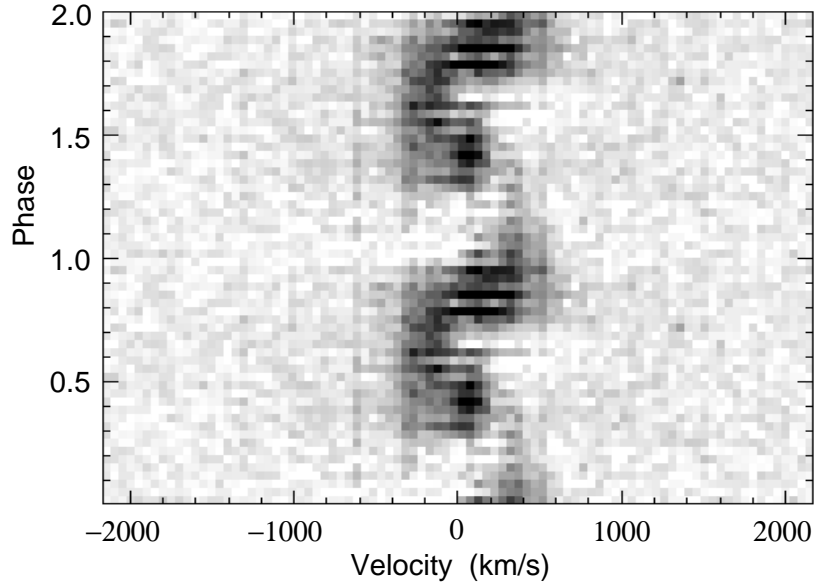}
\caption{The H$\beta$ line of RX\,1712--24 folded on the beat (\beat)
cycle.}
\end{figure}

\section{Curtain-fed accretion}
At this conference I presented `spin-cycle tomograms' of several IPs.
The technique of Doppler tomography can be applied as readily to data folded
on the spin cycle as on the orbital cycle, only the interpretation is
different. Fig.~3 shows tomograms of AO~Psc and PQ~Gem (for a fuller
account see Hellier 1997a, 1999). That of AO~Psc shows the accretion curtain from
the upper pole in the \linebreak ``3 o'clock'' position. Note that it subtends 
\sqig 100\deg\ at the origin, implying that the (line-emitting) accretion 
curtain covers this range of azimuth.  

\begin{figure}[t]       
\vspace*{6.6cm}
\includegraphics{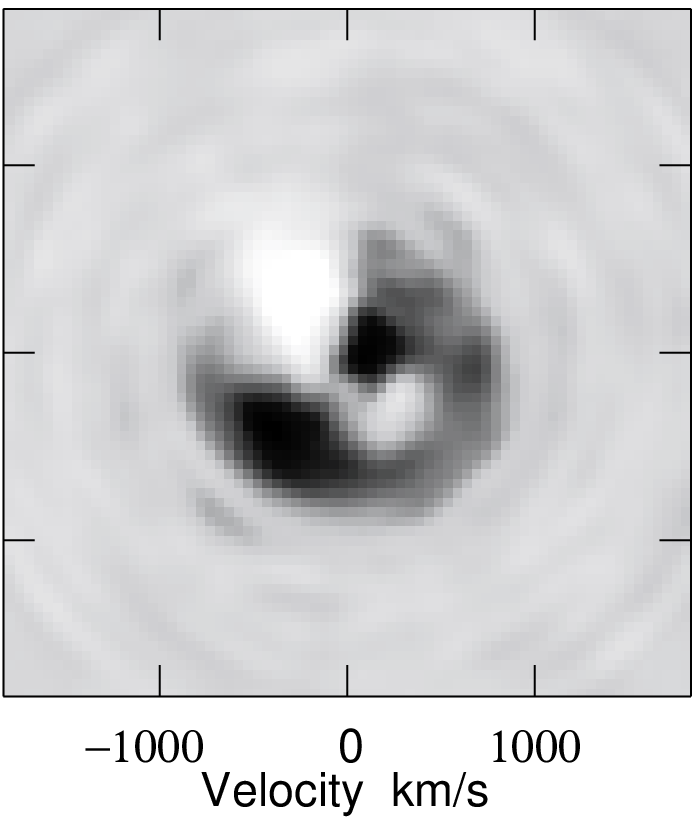}
\includegraphics{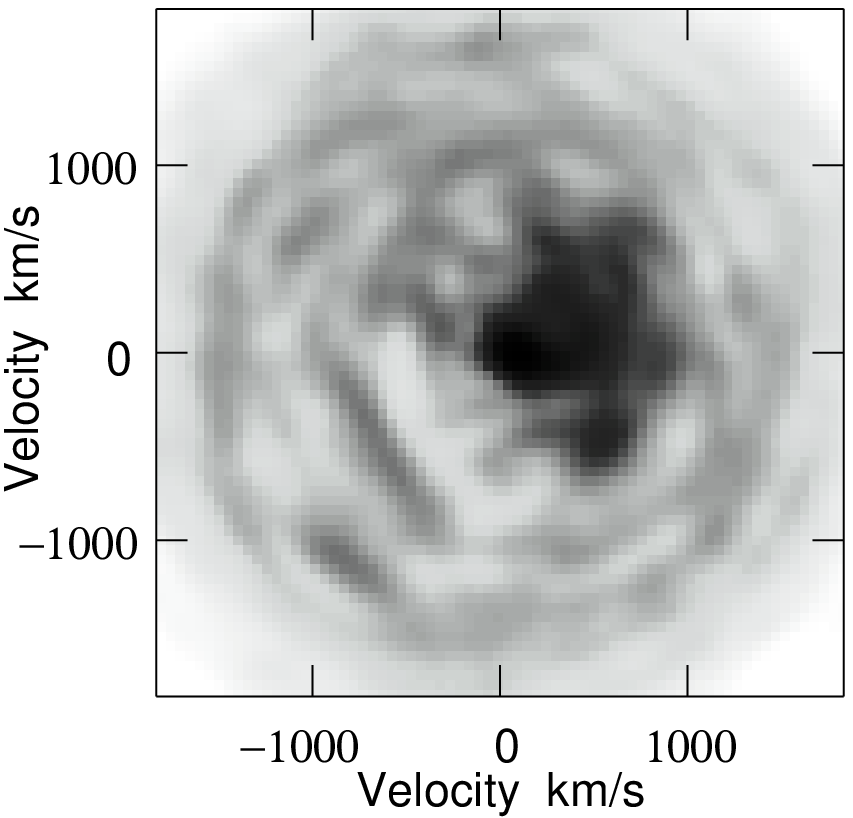}
\caption{Spin-cycle tomograms of the \HeIIl\ lines of AO~Psc (left) and PQ~Gem
(right), from Hellier (1997a; 1999).}
\end{figure}

The PQ~Gem tomogram shows emission from both upper and lower poles. 
In contrast to several IPs, which show maximum blueshift from the upper
curtain when it points away from us (e.g.\ Hellier, Cropper \&\
Mason 1991), the curtains in PQ~Gem appear twisted. If they followed
the expected pattern (maximum blueshift at spin maximum) the poles
would lie symmetrically on the $x$-axis in the tomogram, but they are rotated
by \sqig40\deg. Encouragingly, there
is independent evidence from the X-ray lightcurves (Mason 1997) and from
polarimetry (Potter \etal\ 1997) for the same effect, with material
accreting predominantly along field lines ahead of the magnetic poles.

\subsection{The transition region}
I flag the transition region between the disc and field since it
is the biggest area of uncertainty in IPs. We are still discussing whether
it is an orderly transition from a `normal' disc to magnetically-channeled
infall, or whether the `disc' is a sea of diamagnetic blobs floating in
the magnetic field (e.g.\ Wynn \& King 1995).  The transition region, of
course, determines the accretion footprint.
Given the relatively small accretion areas deduced from observations
(see below) my hunch is that the transition region is relatively abrupt 
and ordered, and that diamagnetic blobs are not able to cross field lines
much. 

Observationally we find it hard enough to measure $r_{\rm mag}$, 
never mind the width of the transition region $\Delta r$. 
We also can't deduce $r_{\rm mag}$ from the magnetic fields
since we have secure $B$ values only for the 3 IPs showing polarised
light [these are PQ~Gem at 9--21\,MG (V\"ath \etal\ 1996; Potter \etal\
1997), RX\,1712--24 at 9--27\,MG (V\"ath 1997), and BG~CMi at 2--6\,MG
(Chanmugam \etal\ 1990)]. 
If someone succeeded in finding $r_{\rm mag}$, though, quoting Patterson (1994), 
``the treasure trove of the world's $\dot{P}$ data, accumulated over decades
and still awaiting a proper interpretation, is theirs for the taking''.
Spin-cycle tomography (see above), offering hints of twisted and
azimuthally-extended curtains, is perhaps the likeliest way forward.

\subsection{The accretion footprint}
If the transition region is hard to pin down, how about the other end
of the curtain, the accretion footprint?  We have fewer observational
constraints than in AM~Her stars. This is because, whereas AM~Hers have
(typically) one dominant accretion region, IPs accrete equally onto two
opposite regions (see below) making the interpretation of their lightcurves
much harder. Recently, though, the deeply-eclipsing IP XY~Ari has
provided secure results. The egress from X-ray eclipse in \sqiglt 2 s 
(Fig.~4) limits the accretion region to \sqiglt 0.002 of the white dwarf area
during quiescence (Hellier 1997b). At the peak of outburst, however, the 
accretion region at the upper pole is visible
at all spin-phases, which implies that it must be extended into an
almost-complete azimuthal ring (Hellier, Mukai \&\ Beardmore 1997). 
There is no constraint on the azimuthal extent during quiescence. 
Thus accretion regions seem to be thin ribbons, covering small areas, 
but extended in magnetic longitude.
This, of course, is what we expect from tracing back field lines from
the disc-disruption region (e.g.\ Rosen 1992), but confirmation
from data is encouraging. 

\begin{figure}[t]       
\vspace*{5.7cm}
\includegraphics{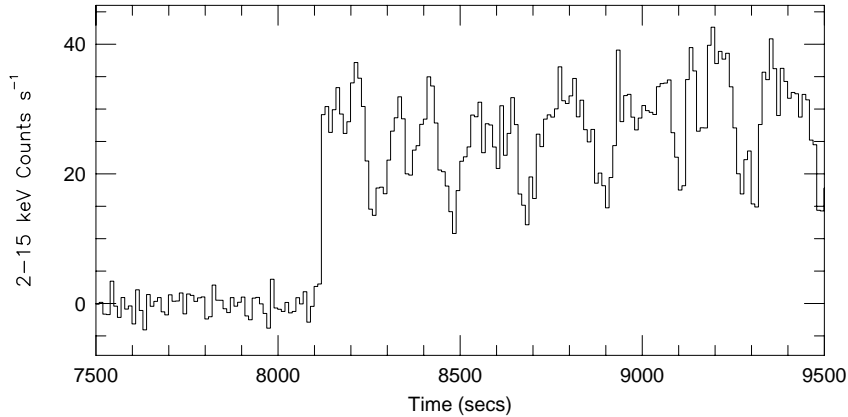}
\caption{XY~Ari emerges from eclipse and begins to pulse, as observed
by {\sl RXTE\/} (from Hellier 1997b).}
\end{figure}

Additional information comes from the black-body-like
soft components in the X-ray spectra of some IPs. Their fluxes lead 
to fractional areas $f$\,\sqig 10$^{-5}$ (Haberl \&\ Motch 1995), and
presumably the hard-X-ray $f$ is smaller still.

\section{X-ray pulse profiles}
The X-ray spin pulses in IPs are potentially the biggest clue to the 
accretion regions on the white dwarf, however their interpretation is still
not settled. Progress is currently hampered by ambiguity in the models, in
that there are many ways of creating a quasi-sinusoid; considerations 
include occultation, photo-electric absorption, electron scattering, 
the effect of non-zero shock heights, and offset or asymmetric dipoles.
Although I don't have space for a comprehensive
account, I can outline the current state of thought. 

Firstly, let me claim that {\it all IPs accrete
roughly equally onto two opposite poles and that both poles contribute
to the pulsation}. The theoretical justification is that we have
never found magnetic monopoles; that a disc feeds both poles roughly
equally (asymmetries have only a minor effect out at $r_{\rm mag}$),
and that even a stream distributes material evenly onto both poles when
averaged over the beat cycle. [In contrast, stream accretion in a
phase-locked AM Her can greatly favour one pole.]\,\ The observational
evidence is the fact that IP hard-X-ray lightcurves never go to zero flux
[as occurs in AM Her stars when only one pole accretes and that pole is on 
the far side of the white dwarf]. Indeed, when the `IP' RX\,1914+24 was 
observed to have zero flux for half its cycle, Cropper \etal\ (1998a)
proposed that it must be an ultra-short-period AM Her star, which appears 
to have been confirmed by further data (Ramsay \etal, this volume).

There is one exception to the above, which is XY~Ari in 
outburst.\footnote{We don't see DQ~Her's white dwarf at all, judging by
the high inclination and lack of X-rays.}
This star has a very high inclination ($>$\,80\deg), a short spin pulse (and
thus a small equilibrium $r_{\rm mag}$), and during outburst the increased 
mass transfer reduced $r_{\rm mag}$ by a factor of two. The combination of
all three allowed the disc to block the line-of-sight to the lower pole 
(Hellier \etal\ 1997), but this exception under extreme
conditions reinforces the point
that normally both poles must be seen.
Many past IP papers have considered only the upper pole, since this is 
conceptually easier, but this simplification is no longer profitable. 

\subsection{Occultation}
Two symmetric, zero-height, 
accretion poles give no net 
modulation through occultation, since one site's appearance compensates
for the disappearance of the site diametrically opposite.  A non-zero 
shock height, though, breaks the symmetry. At large shock heights
($h$\,\sqiggt 0.3) the upper pole is always visible; we see most of 
both poles when the upper pole points away (which I'll call `phase~0' for 
brevity), and at phase 0.5 the lower pole is mostly hidden, so the net 
modulation has a flux minimum at phase 0.5. This may apply to EX~Hya 
(Hellier \etal\ 1991; Allan, Hellier \&\ Beardmore 1998; Mukai, this volume).
At lower, and more typical, shock heights ($h$\,\sqig 0.05) both poles can 
pass over the limb, and occultation produces maxima near phases 0.25 and 
0.75, when both poles are visible because both are on the limb; the result 
is a small-amplitude, double-peaked modulation (Mukai, this volume).
Such features, seen in, e.g., XY~Ari and FO~Aqr, have previously been 
ascribed to offset dipoles (e.g.\ Hellier \etal\ 1997;
Beardmore \etal\ 1998). It will be difficult to separate these 
causes, though Mukai's mechanism is the more natural.

\subsection{Opacity}
If you pour $10^{17}$\,g\,s\up{--1}\ onto $10^{-3}$--$10^{-5}$ of the white dwarf
then a path length of 0.01--0.05\,$R_{\rm wd}$ gives an electron-scattering
optical depth of \sqig 1--100. Photoelectric absorption would 
extinguish soft X-rays 
unless, as expected, the flow is highly ionized. These numbers demonstrate
that opacity in the post-shock flow is highly significant, and is likely
to dominate the pulse profiles, given that occultation largely cancels
out. This is confirmed by the finding that electron-scattering
depths of a few are required to model the X-ray spectra over spin
phase (e.g.\ Hellier \etal\ 1996) and by the observation of 
Compton-broadened K$\alpha$ line profiles in some IPs (Hellier, Mukai \&\ 
Osborne 1998). 

For the high values of $\dot{M}/f$ implied by the above the radial 
optical depth through the accretion shock and the pre-shock flow
will exceed the horizontal optical depth, so that radiation will emerge 
preferentially through the sides of the accretion flow (see Fig.~5).
This is the `accretion curtain' model for producing large quasi-sinusoidal
pulses with both poles acting in phase and with minima when the upper 
pole points towards the observer. It applies most securely to AO~Psc,
whose X-ray lightcurve shows a large-amplitude, sinusoidal pulse which is 
deeper at lower energies, and whose optical emission lines are redshifted 
during the minimum (see Hellier \etal\ 1991, 1996, 
and also see Kim \&\ Beuermann 1995, 1996 for a theoretical model of the 
situation). 

\begin{figure}[t]       
\vspace*{5.2cm}
\includegraphics{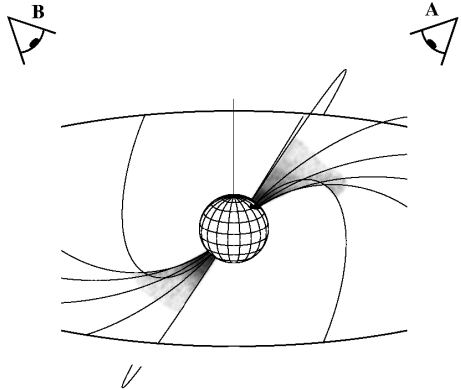}
\caption{\hspace*{-2mm}A schematic accretion column/curtain 
(by A.\,P.~Beardmore).\vspace*{-4mm}}
\end{figure}

This can't be the whole story, however, since many IPs (e.g.\ V405~Aur,
YY~Dra, GK~Per, XY~Ari) show smaller-amplitude, double-humped X-ray pulse profiles,
rather than the `traditional' quasi-sinusoid.  One can perhaps explain
this if the inclination and dipole offset conspire to keep the accretion
curtain from crossing the line-of-sight, allowing occultation effects,
due to non-zero shock heights or asymmetric dipoles, to take over. 
The curtain would never cross the line-of-sight for inclinations 
$i < \delta\!+\!\epsilon$ where $\delta$ is the dipole offset and
$\epsilon$ (18\deg\ for $r_{\rm mag}$\,=\,10\,$R_{\rm wd}$) is the 
magnetic colatitude of accretion.  Many of the IPs showing quasi-sinusoids
(FO~Aqr, AO~Psc, BG~CMi) do seem to be medium-inclination systems 
(as judged by grazing eclipses and/or X-ray orbital dips; Hellier 1995)
where curtain-caused dips would be greatest.  

Another factor is that, as I've pointed out previously (Hellier 1996, see 
also Allan \etal\ 1996; Norton \etal\ 1999), the IPs showing 
double-humped pulses tend to be the faster rotators (e.g.\ V405~Aur,
XY~Ari, YY~Dra, AE~Aqr). Faster rotators will (in equilibrium) have 
smaller values of $r_{\rm mag}$. I suggested that if smaller $r_{\rm mag}$
corresponded to larger $f$ values then the situation above could be reversed,
and optical depths could be lowest radially.
If so, the accretion regions would now act as searchlights, beaming 
X-rays outwards. As the two poles swept into view a double-humped 
pulse would result. This seems to be supported by UV data on the 
fastest rotator, AE~Aqr (e.g.\ Eracleous \etal\ 1994). 

It is unclear whether lowering $r_{\rm mag}$ does lead to a larger $f$:
this would happen if $\Delta r$ was constant as $r_{\rm mag}$ decreased,
but not if it was a fixed fraction of $r$ (the latter would give the same
fractional change in $B$). Since velocities and
therefore velocity dispersions are greater for smaller $r_{\rm mag}$,
though, one might expect a relatively larger $\Delta r$, but the theory of
this topic is sparse.  

One problem with the above idea is that 
one of the stars showing a double-hump, and thus needing a higher $f$, is 
V405~Aur, where the measured
black-body $f$ is only 10$^{-5}$ (Haberl \&\ Motch 1995).  Thus the
picture is unclear and to make progress we need further detailed
investigations of the individual stars.

\subsection{X-ray spectroscopy}
Potentially, X-ray spectra resolved over spin phase should 
sort out the pulse-formation mechanisms. 
However, we are finding that even with the spectral resolution of
{\sl ASCA}, this is not necessarily so.  The reason is that
accretion curtains are found to be patchy, multi-phase absorbers 
complete with electron scattering, giving a much flatter energy
signature than simple photoelectric absorption (e.g.\ Ishida 1991;
Hellier \etal\ 1996). Also, since the bottoms of accretion columns 
occult most readily, and are cooler than the shock, 
occultation can produce the same deeper-at-low-energies 
signature as absorption. In at least two cases [RX\,1238--38
(Hellier \etal\ 1998) and EX~Hya (Allan \etal\ 1998)]
we have been unable to distinguish the two processes with {\sl ASCA\/}
data. Observations with a higher S/N and greater energy range (e.g.\ with {\sl
XMM\/}) will be required. 

\section{Accretion columns} 
We have a good theoretical understanding of the temperature and
density profile of the accretion column for a given white dwarf mass (e.g.\
Aizu 1973), and we have increasingly good X-ray spectra (from {\sl Ginga},
{\sl ASCA\/}, {\sl RXTE} and soon {\sl XMM}). Can we combine the two
to probe accretion columns?

\subsection{White dwarf masses}
In principle we can use a temperature/density profile, together with plasma 
codes, to construct model X-ray spectra, and deduce the 
white dwarf masses by fitting to observations (e.g.\ Cropper, Ramsay \&\ Wu 
1998, see also this volume; Beardmore \&\ Osborne 1999). This is easier 
in IPs than in AM Hers since we can neglect cyclotron cooling, and 
consequently we can assume that electrons and ions have the same 
temperature (e.g.\ Imamura \&\ Durisen 1983).  

There are, though, still considerable uncertainties, including 
clumpiness in the flow; opacity affecting the spectral shape (and
if this includes electron scattering of $\tau$\,\sqig 1 
it will affect all energies); the contribution of X-rays reflected 
by the white dwarf surface; the uncertainty in $\dot{M}/f$ (which sets
the shock height); uncertainty in the shape of the accretion
region (which affects path lengths); the need for a low-temperature
cutoff where the column goes optically thick as it merges with the
white dwarf; and the mass-radius relation of a hot white dwarf 
possessing an accreted hydrogen envelope. 

\begin{figure}[t]             
\vspace*{7.5cm}
\includegraphics{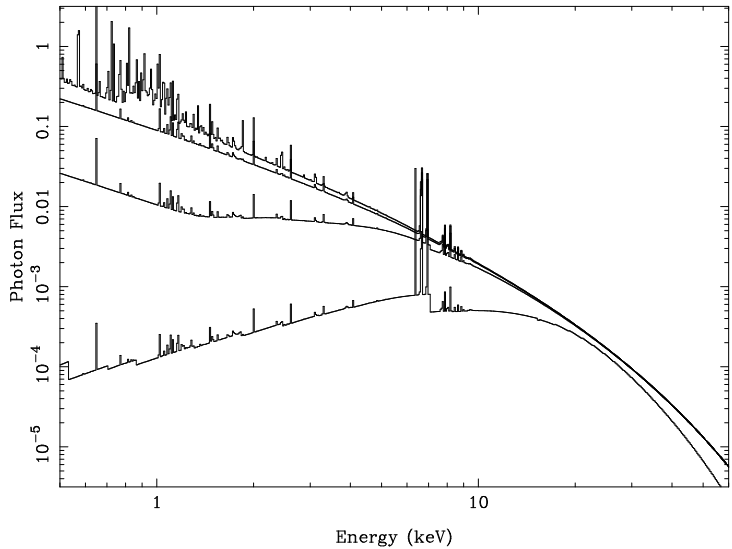}
\caption{Model accretion column spectra using the 
{\sc mekal} plasma code and an Aizu distribution of temperatures (uppermost).
The next curve has a low-energy cutoff at 2 \kev, as 
though the column goes optically thick. The third curve additionally 
includes a complex absorber, as required to model the {\sl ASCA\/}
spectra of AO~Psc. The bottom curve is the component
reflected from the white dwarf. (By A.\,P.~Beardmore)}
\end{figure}

Given this it is perhaps not surprising that there are still discrepancies
between masses derived by this method and those from other methods (the 
X-ray spectra tend to give higher masses --- see Ramsay 
\etal\ 1998; Cropper \etal, this volume). However, the estimates by
different methods are currently converging, and with some more tuning
the X-ray spectra could turn into powerful tools for deriving 
the white dwarf masses of a whole class. 

One factor I haven't seen discussed is that most treatments
assume that the accretion flow falls into the white dwarf from infinity.
In an AM~Her, with quasi-radial infall from the $L1$ point, this is
a fair assumption, but in an IP, assuming disruption of the disc at
$r_{\rm mag}$, it is unclear how much, in any, of the Keplerian velocity
translates into speed down field lines. If we assume freefall only from
$r_{\rm mag}$\,\sqig 10\,$R_{\rm wd}$ then the kinetic energy at the
white dwarf is reduced by 10\%. Further, the enforced co-rotation with
rapidly-spinning field lines will produce a centrifugal force which further
slows the infall. A simple calculation, including a centrifugal force term 
in the freefall, shows that for a fast rotator (in
equilibrium at $r_{\rm mag}$\,\sqig 7\,$R_{\rm wd}$) the two effects
combine to a \sqig 21\%\ reduction in kinetic energy. This is, for
example, sufficient to remove the remaining discrepancy between 
estimates of XY~Ari's white dwarf mass by this method compared to methods
using its eclipse (see Cropper \etal, this volume). 

\subsection{Broadened lines}
\begin{figure}[t]             
\vspace*{8.0cm}
\includegraphics{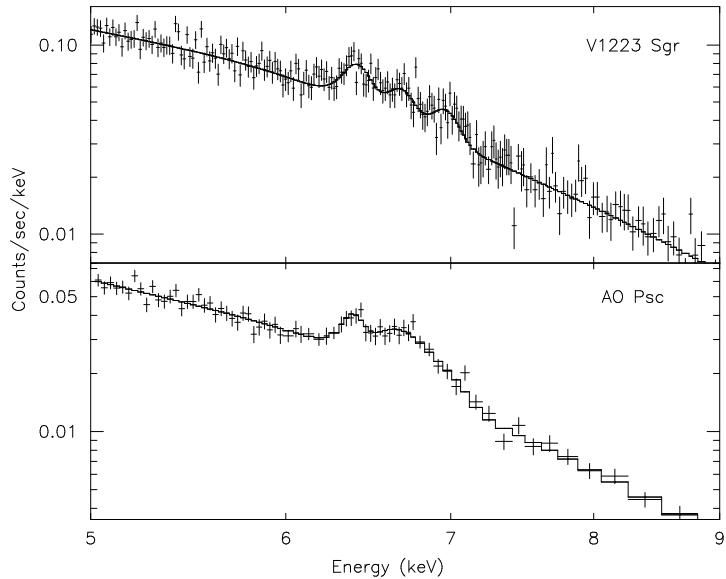}
\caption{{\sl ASCA\/} spectra 
showing that the components of the K$\alpha$ line (6.41, 6.70 \&\
6.97\,\kev) are narrow in V1223~Sgr but that the thermal lines (6.70 \&\
6.97\,\kev) are broadened in AO~Psc.}
\end{figure}
Another potential probe of the accretion column occurs through the
discovery of broadened iron K$\alpha$ lines in the X-ray spectra
of some IPs (Hellier \etal\ 1998; Fig.~7). We showed that 
Doppler broadening was an insufficient explanation, and concluded
that Compton scattering in the post-shock region was broadening the
lines. We suggested the picture drawn schematically in Fig.~8, where
the hot, optically-thin, upper region radiates little line emission.
In the dense, optically-thick, lower region, 
line photons are destroyed by multiple scatterings.
In the transition region, however, resonant trapping of K$\alpha$
photons allows them to be Compton-scattered once and only once as they
emerge. The degree of broadening is thus a probe of the column temperature
at the transition to optical thickness.  This could explain why some 
IPs show broad lines while others don't: if the transition to optical
thickness occurs at temperatures too low for significant iron K$\alpha$
emission then only narrow lines will be seen. 

\begin{figure}[t]             
\vspace*{7.5cm}
\includegraphics{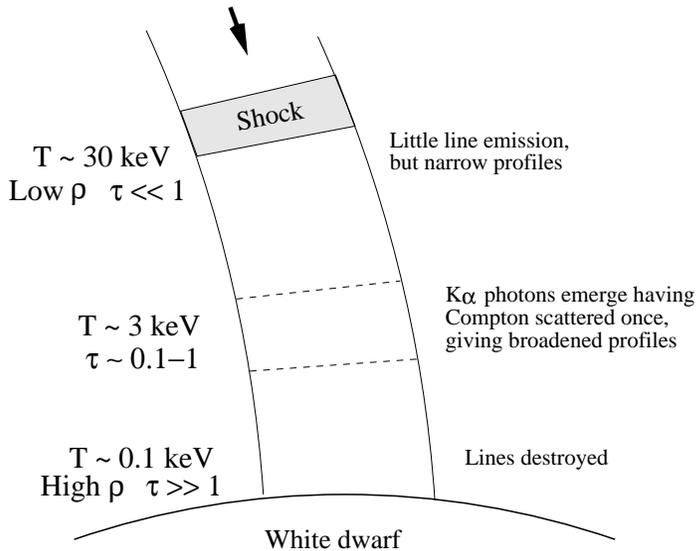}
\caption{A schematic view of iron K$\alpha$ emission from a column.}
\end{figure}

\subsection{Soft IPs}
Owing to {\sl ROSAT\/} we know of three IPs (PQ~Gem, V405~Aur \&\ 
UU~Col; Mason \etal\ 1992; Haberl \etal\ 1994; Burwitz \etal\ 1996)
that show soft black-body components similar to those in AM Her stars
(a fourth star, RX\,1914+24, I now class as an AM~Her, see Cropper \etal\
1998a). 

People have often supposed that IPs have larger $f$ values 
than AM~Hers, since IPs accrete from a disc over a range of azimuth, 
whereas AM~Hers accrete from a narrow stream. 
The soft components in IPs, however, are convincing me otherwise. 
Firstly, they have measured temperatures (38--57 \ev) in the range that
AM~Hers do, whereas larger areas would lead to lower temperatures.
Secondly, the areas derived from their fluxes (e.g.\ 10$^{-5}$\,for
V405~Aur at 300\,pc; Haberl \&\ Motch 1995) are no higher than those
found for AM~Hers (see Stockman 1995). 

Why is this?  In an AM~Her,
the radial infall of the stream helps it to punch a hole in the
magnetosphere. It breaks up into blobs which become magnetically
controlled at different field strengths (and hence different radii) 
depending on their size and density. Thus the ballistic-to-magnetic 
transition occurs over a large volume, corresponding
to a large footprint on the white dwarf. This appears to have been
observed in spectroscopy of the accretion stream (Sohl \&\ Wynn, this
volume). In contrast, passage through a disc in an IP will 
destroy blobs, and during disc-disruption the tangential
Keplerian velocity won't help the material to penetrate the field.  The field
lines are, presumably, scooping up material at a near-constant rate as
they rotate, and since diffusion times are longer than spin cycles,
this is likely to occur at a constant radius. Although the theory of 
this is uncertain, it could lead to a homogenous, rather than blobby,
accretion flow.  The small $\Delta r$ would map to a smaller footprint on 
the white dwarf, compensating for the larger range of azimuth. 

Two further observations support this idea. Firstly, AM~Hers can show
high-amplitude, erratic, soft-X-ray lightcurves, as individual blobs hit the
photosphere (Heise \etal\ 1985); nothing as dramatic has been seen in IPs.
Secondly, whereas the soft component is widespread in AM~Hers, and thought
to be mainly due to blobs penetrating the photosphere and thermalising,
only 3 of 23 IPs show comparable soft components, suggesting a general
absence of blobby accretion. 

So why do those 3 show soft components?  We expect some soft emission
from irradiated areas around the accretion region in all IPs, but what 
does it correlate with? 
With only 3 systems this is hard to answer, but it doesn't seem to
be field strength: PQ~Gem and RX\,1712--24 both have 10--20\,MG fields
but the former shows a soft component and the latter doesn't. 
Nor does it appear to be absorption: that to RX\,1712--24 is no higher 
than that to V405~Aur (Motch \&\ Haberl 1995).  Inclination? Tricky to
say, since inclination estimates are only reliable for eclipsing systems.
Accretion rate? These estimates are always unreliable!  
Could the heated regions be veiled by the accretion curtain? 
Perhaps, but in simple pictures the column-base at one of the poles
should always be seen at some spin phase. Thus, this topic remains
unsolved. 

\section{Outbursts}
Despite the fact that, even up to 1998, authors have stated that outbursts
are discordant with an IP classification, at least 6 confirmed IPs have 
shown outbursts (e.g.\ Warner 1996; Hellier \etal\ 1997).
I want to publicise the fact that there appear to be two types of outburst.
YY~Dra, GK~Per \&\ XY~Ari all show `normal' dwarf-nova-like outbursts,
with any differences being explained by their occurrence in a truncated
disc (e.g.\ Angelini \&\ Verbunt 1989).  The other three, EX~Hya,
TV~Col \&\ V1223~Sgr, show shorter, lower-amplitude outbursts with a
range of observational properties at odds with normal behaviour 
(Hellier \etal\ 1997 and references therein). It appears that the normal
instability is suppressed and replaced by a different instability 
(see Warner 1996). We don't know what this is, though secondary-star
instabilities have been suggested, but it remains a topic in need of
attention by theorists. Can the outbursts in these 3 stars be explained
within the standard disc-instability paradigm?

\end{document}